\documentclass[pra,amsmath,amssymb, twocolumn, showpacs]{revtex4}

\usepackage{dcolumn}

\usepackage{bm}

\usepackage[dvips]{graphicx}

\usepackage{epsfig}

\begin{document}

\title{Three bodies bind even when two do not: Efimov states and Fano resonances in atoms and nuclei}

\author{A.\  R.\  P. Rau$^{*}$}
\affiliation{Department of Physics and Astronomy, Louisiana State University,
Baton Rouge, Louisiana 70803-4001, USA}


\begin{abstract}

Efimov's prediction more than three decades ago that three-body bound states can exist when the pairwise attractions do not bind or only support weakly bound states of a pair, has remained unconfirmed till just the past year. This lecture provides the pedagogical background for recent work on Efimov states in neutron-rich nuclei done with I. Mazumdar (TIFR) and V. S. Bhasin (Delhi University), and published in Phys.\ Rev.\ Lett.\ {\bf 97}, 062503 (2006). Both these nuclear systems and recent observations of cold cesium atoms provide the first clear evidence for the existence of Efimov states.  

\end{abstract}

\pacs{21.45.+v, 42.50.Ct, 03.65.-w, 32.80.Dz}

\maketitle

\section{Introduction}

A romantic triangle has been one of the staples of theatre and cinema, that the appearance of a third party can change qualitatively the relationship between a pair. Curiously, quantum physics also has something analogous as was first pointed out
over three decades ago by Efimov \cite{ref1}. Under fairly general conditions with only pairwise interactions, a three-body system can support bound states when none of the three pairs constituting it is bound or one or two pairs are barely bound. Indeed, there may even exist an infinity of three-body bound states! But, in spite of many attempts over the years, it is only within the last year that a first experimental observation of Efimov states has been reported, in ultracold cesium trimers \cite{ref2}. And, simultaneously and independently, we \cite{ref3} have presented a study of neutron-rich nuclei wherein characteristic asymmetric resonance profiles as described by Fano \cite{ref4} may provide a diagnostic for such Efimov states. This lecture provides the various elements of physics that underlie these phenomena to provide a pedagogical understanding. 
 
Several themes come together in explaining the Efimov effect and the asymmetric resonances that it may lead to. These are: the nature of quantum-mechanical binding and the role played by $1/r^2$ potentials, dipole-bound states, resonance profiles and especially their asymmetric shape, scattering length and its tuning. We will consider each in turn.

\section{Quantum binding and $1/r^2$ potentials}

From its very beginnings, non-relativistic quantum mechanics explains the stability of the Bohr atom and thereby of all matter (the Pauli principle for electrons as fermions also needs to be invoked) as a balance between the attractive Coulomb potential $-(e^2/r)$ and the quantum  kinetic energy $(\hbar^2/2mr^2)$. The balance provides the scale of atomic sizes and energies. A part of the kinetic energy, that arising from the angular motion, is also familiar as the angular momentum ``potential", $\ell(\ell+1)\hbar^2/2mr^2)$. Many-particle systems, when viewed in higher-dimensional hyperspherical coordinates \cite{ref5,ref6}, have similar expressions in terms of a hyperspherical radial variable $R$ and the total orbital angular momentum $L$. A central feature that follows on dimensional grounds alone is that with $\hbar$, a reduced mass, and a length, the combination $(\hbar^2/MR^2)$ is an energy.

The same balance also leads to the conclusion that an attractive $1/r^2$ potential marks the dividing line between a spectrum with an infinite and one with at most a finite number of bound states. Because of the quadratic scaling of the radial kinetic energy, any potential that falls off at large distances slower than $1/r^2$ will support arbitrarily weakly bound states and thus an infinite number. By spreading out the wave function over a distance $\Delta r$, with $\Delta r/r$ held at some small constant value, the kinetic energy can be made smaller relative to the potential to get net binding. Potentials that fall off faster than $1/r^2$ can only have a finite number of bound states. For a $1/r^2$ potential itself, with both terms scaling in the same manner, details of the strength of the attraction matter. Above a certain critical strength, the potential will support an infinite number, and below that only a finite number of bound states.

\section{Dipole bound states}

An attractive dipole potential, $-a/2r^2$, can be described in the above terms of an angular momentum potential through a complex angular momentum, $(-\frac{1}{2}+i\alpha)$, with $\alpha=\sqrt{a-\frac{1}{4}}$. Many polar molecules and the hydrogen atom in its excited states, because of the peculiar degeneracy of opposite parity states, present such attractive dipole potentials to an external electron. The condition for bound states of energy $\epsilon =-\kappa^2/2$ (in atomic units) is given by (Sec.\ 5.6 of \cite{ref5})

\begin{equation}
\alpha \rm{ln} (2/\kappa) - \rm{arg} \Gamma (1-i\alpha) = (n+1/2)\pi,
\label{eqn1}
\end{equation}
which leads to a characteristic infinite sequence that piles up exponentially at threshold:

\begin{equation}
\epsilon_n = \epsilon_0 e^{-2n\pi/\alpha}.
\label{eqn2}
\end{equation}
Fig.\ 1.5 of \cite{ref5} gives an example of one of these resonances observed in the electron + H($n=2$) system, and Fig.\ 10.16 the corresponding potential well in a hyperspherical description of the six coordinates of the two electrons. Note the very asymmetric profile of this resonance in the well marked as minus which has attractive $1/r^2$ behaviour at large $r$. The exponential dependence on $n$ in Eq.~(\ref{eqn2}) means a very ``compressed" spectrum below threshold, with the larger $n$ states so close to it that higher members of an infinite sequence are rarely observed.   

\section{Resonances as interference phenomena}

Resonances are ubiquitous in physics, from mechanical and electromagnetic oscillators in classical physics to various quantum systems in all sub-areas of physics: atomic, condensed matter, nuclear and particle. Most often, the resonant response or cross-section is expressed by a symmetric Lorentzian, or Breit-Wigner, profile,

\begin{equation}
\sigma = \frac{A}{(E-E_r)^2+(\Gamma/2)^2} = \frac{\sigma_0}{1+\epsilon^2},
\label{eqn3}
\end{equation}
with $E_r$ the energy position of the resonance, $\Gamma$ its width, and $\epsilon \equiv (E-E_r)/(\Gamma/2)$ a reduced energy measured with respect to $E_r$ in units of the half-width.

In essence, however, all resonances result from an interference between alternative pathways for connecting an initial and a final state just as in the canonical two-slit example. As with any quantum interference phenomenon, the cross-section in general displays both constructive and destructive interference in traversing a resonance. Thus, as shown in Fig.\ 1(a), in the H$^-$ or its equivalent He system, doubly excited states such as $2s^2$ lie embedded in the one-electron continuum $1sEs$, so that the same final continuum energy state can be reached from some lower level either directly or through an alternative pathway that goes through the embedded discrete configuration. The resulting resonance, in say elastic scattering of electrons from H or He$^+$ in that energy range, will show the effects of interference through an asymmetric profile, first described in this very context by Fano \cite{ref4}:

\begin{equation}
\sigma = \sigma_0 \frac{(q+\epsilon)^2}{(1+\epsilon^2)}.
\label{eqn4}
\end{equation}
In contrast to Eq.~(\ref{eqn3}), this expression involves a third parameter, the Fano profile index $q$, besides $E_r$ and $\Gamma$. The cross-section expression above is in general asymmetric, depending on the value of $q$, reducing to the symmetric Lorentzian in Eq.~(\ref{eqn3}) for $q = \infty$ and 0. See Figs. 1.5, 8.1, and 8.3 of \cite{ref5} for various examples. $q$ expresses the ratio of the two amplitudes involved in the interference, so that the reduction to Lorentzian is when one pathway dominates, as often happens in nuclear and elementary particle physics wherein the background is small in the vicinity of the embedded discrete state. But, in atomic, molecular, and many condensed matter systems \cite{ref7}, other values of $q$ and asymmetric profiles occur quite commonly.   

\begin{figure}
\includegraphics[width=3in]{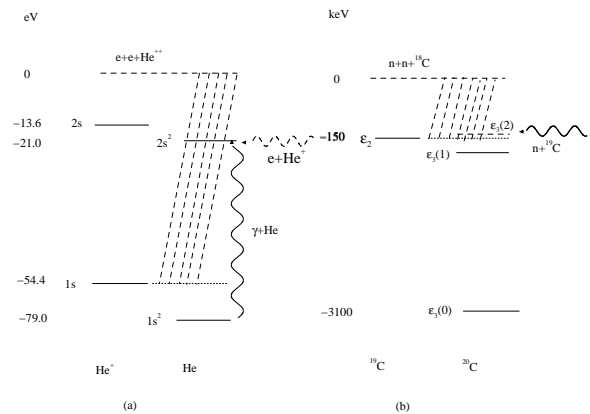}
\caption{
Comparison between He and $^{20}C$ as three-body systems in atoms and nuclei. (a) Ground and first excited state (there are infinitely many) of He$^+$ and the lowest bound states of He (again there are infinitely many) attached to them are shown, along with their energies below the fully dissociated limit of two electrons and the helium nucleus. Cross-hatched region is the ionization continuum of He built on the ground state of He$^+$, that is, the two-electron states $1sEs$, $E$ being the energy of the continuum electron. The $2s^2$ state lies embedded in this continuum, that is, is degenerate with $1sEs$ and, therefore, mixes with it to give the quasi-bound resonance state. The resonance can be accessed either by (e + He$^+$) scattering or by photoexcitation from the ground state as shown (because of dipole selection rules, a single photon will reach similar states of $2s2p \,^{1}P^o$ symmetry whereas two-photon absorption would be necessary to reach the $2s^2$). (b) Similar schematic for the nuclear system. Now $^{19}$C has only one bound state, its ground state $\epsilon_2$. For an appropriate value of $\epsilon_2$, one bound Efimov state $\epsilon_3 (1)$ and a second which just fails to be bound, $\epsilon_3 (2)$, are shown, together with the ground state of $^{20}$C. The state $\epsilon_3 (2)$, lying embedded in the n+$^{19}$C continuum, manifests as a resonance in low energy elastic scattering of n+$^{19}$C. For somewhat more negative $\epsilon_2$, the first Efimov state $\epsilon_3 (1)$ also fails to be bound and will appear as a resonance in n+$^{19}$C scattering as in Fig.\ 2. From \cite{ref3}}. 
\end{figure}
 
\section{Scattering length and its tuning}

In discussing near zero-energy states, whether weakly bound or in the low energy continuum, a single parameter suffices to characterize the $s$-wave radial wave function, the scattering length. Whether for the van der Waals potentials between two cold atoms or potentials between nucleons, let us model them with a short range potential and consider the radial wave function just outside the potential well. With $\ell$, $V$, and $E$ all zero or nearly so, the radial equation contains just the second derivative in $r$ so that the wave function reduces to only constant and linear terms and is proportional to $(1-r/a)$, where $a$ is the scattering length; see Fig.\ 4.3 of \cite{ref5} or Fig.\ IX.9 of \cite{ref8} for a figure popularized by Fermi. In an attractive potential, a positive scattering length corresponds to the possible existence of a bound state because the downward linearly sloping function can connect to an exponentially decaying behaviour in addition to the possibility of connecting to oscillating behaviour appropriate to scattering states. On the other hand, a negative scattering length, with its attendant upwardly sloping radial function, can only connect to the oscillations of a continuum function and no bound state exists in this case. A very large magnitude of the scattering length provides the dividing line between the two cases, $a$ large and negative just falling short of binding and $a$ large and positive pointing to a weakly bound state.

The historical example of the nucleon-nucleon system is very illustrative, the only bound state of this system being of neutron-proton with triplet spin. This state, the deuteron, is weakly bound (2.2 MeV, to be contrasted with the canonical value of 8 MeV for the average binding energy of  a nucleon in any nucleus), corresponding to a fairly large, positive, value of the scattering length $a_t$. In the singlet sector of neutron-proton, the absence of any bound state got ascribed to a negative scattering length $a_s$, which in magnitude is even larger. Thus, the attractive potential between a neutron and a proton is significantly spin dependent, the singlet sector just failing to bind. On the other hand, low energy neutron-proton scattering cross-section, which depends only on the square of the scattering length, is dominated by the singlet configuration. The singlet attraction falls just short of binding, the triplet suffices just to bind one and only one state.

In recent years, in the study of ultracold atoms and Bose-Einstein condensates (BEC), the ability to control the scattering length describing the interaction between atoms at large distances has played a central role. Because an external magnetic field can change the Zeeman energies of fine structure states and thereby the atom-atom potential wells, one can use the magnetic field as a knob under the experimenter's control to change the strength of that interaction continuously over some range, so much so that one can go from large negative to large positive values of $a$ \cite{ref9}. This ``magnetic tuning" has been used very effectively, especially in studies of sudden collapse of condensates (when the scattering length's sign is changed), formation of molecular condensates, and studies of BEC-BCS (Bardeen-Cooper-Schrieffer) crossover \cite{ref10}.

\section{The Efimov effect}

We can now put together the themes developed in the above sections to understand the Efimov effect \cite{ref1,ref11}. On dimensional grounds alone, if no other length scales are available, the overall size of the system provided, as for instance, by the hyperspherical radius $R$ (given by the square-root of the sum of the squares of the two lengths of the three-body system in its centre of momentum frame) is the only length, and thereby $(\hbar^2/MR^2)$ the only combination with dimensions of energy. This is precisely the situation for most $R$ once outside the range $r_0$ of the pairwise interaction when $|a| \rightarrow \infty$. Thus, the effective potential in the system is an attractive $1/R^2$ potential. One can picture two bodies within the range of their interaction with the third distant from both, and consider small changes in their positions. The slight change in the first separation being dwarfed in the overall $R$, the resulting slight change in attraction is to first order proportional to such a $1/R^2$ behaviour. With no other repulsive $1/R^2$ potentials that may overwhelm, there will then appear an attractive $1/R^2$ potential for the three-body system. The Efimov effect occurs, therefore, only for states of zero total angular momentum $L$.              

Thus, only when $|a|$ is large, so that the two-body binding is either weak or absent, are Efimov states formed. As shown by Efimov, the number of three-body bound states in the $1/R^2$ potential is $(1/\pi)\ln(|a|/r_0)$ and infinite in the $|a| \rightarrow \infty$ limit, otherwise finite. As the two-body binding increases, with $a$ itself providing another length, the $1/R^2$ attraction disappears. This provides yet another astonishing element of the phenomenon, that the stronger is the pairwise attraction and pair binding the fewer are the Efimov three-body states, while the weaker that attraction so that two body binding is discouraged, the larger is the number of three-body bound states.

Recent observations of an Efimov state in cesium trimers indeed occurred through tuning of the scattering length \cite{ref2,ref12}. By magnetically tuning across a resonance, the cesium-cesium interaction's $a$ was continuously changed from -2500$a_0$ to 1600$a_0$, where $a_0$ is the Bohr radius. Formation of cesium trimer Efimov states around -800$a_0$ was monitored in terms of escape from the trap of 10 nK cold cesium atoms; see figures in \cite{ref2}. Since the magnetic field was used for tuning $a$, these experimentalists made a purely optical trap for cold cesium and did not use magneto-optical traps as do most groups working with BEC.

At the same time, in a completely different system and of very different energies, we studied very neutron-rich nuclei such as $^{19}$B and $^{20}$C. $^{18}$C is bound but, in part because of the well-known difference between odd and even nuclei, $^{19}$C is only weakly bound at best as shown in Fig. 1(b), with a binding energy of about 100 keV. With the two-neutron system also falling just short of binding, the total $^{20}$C system as n+n+$^{18}$C satisfies the conditions for the Efimov effect. Using the two-body $^{19}$C binding as a parameter, we calculate resonances of $^{20}$C due to Efimov states which lie embedded in the n+$^{19}$C continuum as shown in Fig.\ 1(b). In the figure, a two-body binding energy of 150 keV shows two of these Efimov states, the higher one just above that continuum threshold. Exactly analogous to the atomic doubly-excited states shown in Fig.\ 1(a) alongside and discussed earlier, such a state will manifest itself as a resonance, with interference between the two pathways to the continuum, one through the Efimov state and one direct into the underlying continuum. A slightly larger two-body binding of 250 keV will move the lower state, $\epsilon_3(1)$ just above the threshold and only that one Efimov state will appear as a resonance. The elastic cross-section shown in Fig.\ 2 displays this resonance. Note immediately its non-Lorentzian, asymmetric shape. We have fitted it to a Fano profile as in Eq.~(\ref{eqn4}) with the parameters shown in the caption, especially a profile index, $q=4$. As already noted, resonances in nuclear physics are generally symmetric Lorentzians but it is precisely in the context of an Efimov state embedded in a continuum that asymmetry should be expected. The very loosely bound Efimov state with large spatial extent overlaps significantly with a low energy continuum state, so that the two pathways are more on par and both the constructive and destructive interference of the Fano resonance clearly seen. 
   
\begin{figure}
\includegraphics[width=3in]{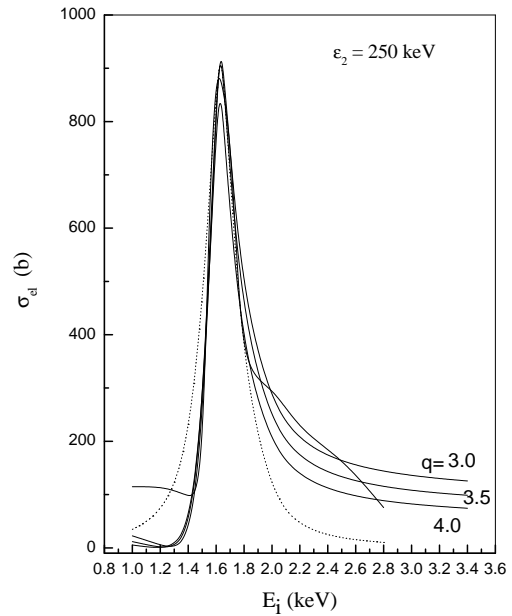}
\caption{
Elastic cross section for n-$^{19}$C scattering vs. centre-of-mass energy for a n-$^{18}$C binding energy of 250 keV. Calculated cross section (full curve) is fitted to the resonance formula in Eq.(4), the best fit obtained for parameters $E_r$ = 1.63 keV, $\Gamma$ = 0.25 keV, and $q$ = 4.0. The dotted line represents a Breit-Wigner fit to the calculated curve. From \cite{ref3}.}
\end{figure}

\section{Conclusions and further aspects}

With the advent of special purpose machines built for creating very neutron-rich nuclei such as $^{20}$C, we can anticipate experimental observation, either through elastic scattering or in fragmentation on a heavy target. In other systems such as cold atoms, it will also be interesting to see more direct evidence for Efimov states. Observing a sequence of them that fit the exponential law of Eq.~(\ref{eqn2})
would provide unambiguous confirmation but may be difficult. Since $\exp(-2\pi)$ is, approximately, 1/500, the second member of the sequence will, for realistic values of $\alpha$, be closer to threshold by that factor as compared to the lowest member and thereby difficult to observe.

Efimov's work and most others that followed considered spinless bosons. Very recently, the effect of quantum statistics for identical particles has been investigated \cite{ref13}. Given the nature of these states, that they are weakly bound and of large extent, with all three particles far apart, the boson or fermion nature would seem to be unimportant, except in ruling out certain values of $L$. As already noted, the weak attractive $1/R^2$ will be overwhelmed by any non-zero $L(L+1)/R^2$, so that any three identical fermion system for which $L=0$ is forbidden by the Pauli principle will not support Efimov states. Apart from this restriction on $(S,L,J)$ due to statistics of identical particles, one could expect Efimov states to be largely insensitive to whether we deal with bosons or fermions. In our own study, $^{20}$C has two fermions and one boson, whereas $^{19}$B has all three fermions, two of them identical.

Finally, it is worth considering whether similar states exist for more than three particles. Some studies have claimed the absence of the Efimov phenomenon for four or more particles \cite{ref14}. A simple case can be made for this in a hyperspherical analysis. A  kinematic effect of eliminating the linear derivative term in $(1/R)(d/dR)$ introduces in general an effective $1/R^2$ potential in the Schr\"{o}dinger equation \cite{ref5}. When only two radial distances are involved as in the three-body problem, this term is attractive, for higher number of particles repulsive. Again, any such repulsion will overwhelm the weak $1/R^2$ attraction underlying the Efimov states \cite{ref15}. This would argue for Efimov states in three, and possibly four (when the above kinematic $1/R^2$ vanishes), particles but not beyond that. Curiously, to return to the sociological analog at the beginning of this lecture, the romantic triangle is pervasive in literature but entanglement of four or more persons is of less interest and import!

This work has been supported by the U.S. National Science Foundation Grant 0243473 and by the Roy P. Daniels Professorship at LSU. I also thank the Tata Institute of Fundamental Research for its hospitality and support as an adjunct professor during the course of this work.


\begin{thebibliography}{}
\bibitem{ref15} The occurrence of the Efimov effect for three particles is thereby related to the system's two-dimensional nature, that three points define a plane. It is only in the two-dimensional Laplacian that elimination of the $(1/R)(d/dR)$ term gives an attractive $1/R^2$ term, the counterpart for four or more dimensions being repulsive \cite{ref5,ref6}.
\end{thebibliography}
\end{document}